# Tunable Polarization-Multiplexed Achromatic Dielectric Metalens

Xiangnian Ou[1], Tibin Zeng[2], Yi Zhang[1], Yuting Jiang[1], Zhongwei Gong[1], Fan Fan[2], Honghui Jia[1], Huigao Duan[1,4] and Yueqiang Hu[1,3*]

[1] National Research Center for High-Efficiency Grinding, College of Mechanical and Vehicle Engineering, Hunan University, Changsha 410082, China

[2] Key Laboratory for Micro/Nano Optoelectronic Devices of Ministry of Education & Hunan Provincial Key Laboratory of Low-Dimensional Structural Physics and Devices, School of Physics and Electronics, Hunan University, Changsha 410082, China

[3] Advanced Manufacturing Laboratory of Micro-Nano Optical Devices, Shenzhen Research Institute, Hunan University, Shenzhen 518000, China

[4] Greater Bay Area Institute for Innovation, Hunan University, Guangzhou 511300, China

[*] Corresponding authors. Email: huyq@hnu.edu.cn

**Abstract:**

Tunable metasurfaces provide a compact and efficient strategy for optical components that require active wavefront shaping. Varifocal metalens is one of the most important applications. However, the existing tunable metalens rarely serves broadband wavelengths restricting their applications in broadband imaging and color display due to chromatic aberration. Herein, we demonstrate an electrically tunable polarization-multiplexed varifocal achromatic dielectric metalens integrated with twisted nematic liquid crystals (TNLCs) in the visible region. The phase profiles at different wavelengths under two orthogonal polarization channels are customized by the particle swarm optimization algorithm and optimally matched with the metaunits database to achieve polarization-multiplexed dispersion manipulation including achromatic performance. By combining the broadband linear polarization conversion ability of TNLC, the tunability of varifocal achromatic metalens is realized by applying different voltages. Further, the electrically tunable customized dispersion-manipulated metalens and switchable color metaholograms are demonstrated. The proposed devices will accelerate the application of metasurfaces in broadband zoom imaging, AR/VR displays, and spectral detection.

**Introduction**

Metasurfaces are planar optical elements composed of scatterers with subwavelength sizes arranged on a two-dimensional plane. By adjusting the shape, size, position and orientation of the scatterers, almost arbitrary control of the phase[1], amplitude[2], polarization[3] and frequency[4] of light can be achieved. With such ability of multi-parameter modulation, large-volume single-function refractive optical elements can be designed into ultrathin and multifunctional elements[5]. Various metasurface-based devices have been demonstrated, such as metalenses[6, 7], metaholograms[8-10], vectorial beam generators[11, 12], structural color[4, 13] and nonlinear optical metasurfaces[14, 15]. Due to the strong demand for integrated and miniature optical components in zoom imaging, optical displays and light detection and ranging (LiDAR) systems, actively tunable metasurfaces have attracted extensive attention and research[16]. Various tuning mechanisms have been proposed, such as the use of phase-change materials[17, 18], mechanical actuation[19, 20], free carrier effects[21, 22], chemical reactions[23] and surrounding environment manipulation[24-26]. Particularly, liquid crystals (LCs) are widely used to implement tunable metasurfaces due to their unique birefringence properties and mature control mechanisms[25, 27].

Varifocal metalens is one of the important applications of tunable metasurfaces and is expected to replace bulky refractive optical elements. However, most tunable metalenses operate at a single wavelength, and dispersion limits their practical applications in fields such as broadband imaging and color displays[26, 28]. Dispersion is a fundamental property of materials, which causes chromatic aberration in imaging.

The conventional dispersion manipulation requires the superposition of several components[29], resulting in a bulky and complex system. The units of metasurfaces have a higher degree of design freedom[30-32], which opens up the possibility of using a single lightweight metasurfaces to control dispersion. So far, some strategies for multi-wavelength or broadband metasurfaces have been proved by using multi-layer metasurfaces[33], interleaved structures[34] and phase compensation[35, 36]. The phase compensation method can achieve achromatic performance with a non-interleaved metaunit, but it rarely combines with other parameters (e.g., polarization) to achieve multi-channel dispersion manipulation. In addition, the method of combining dispersion manipulation with existing tunable mechanisms is still to be explored.

In this work, we demonstrate an electrically tunable polarization-multiplexed achromatic dielectric metalens (TPAM) integrated with twisted nematic liquid crystals (TNLCs) in the visible region. Vertical stacking of metalens and TNLCs offers a simpler device preparation process and control mechanisms. A single non-interleaved rectangular composite structure is used as a metaunit and its phase response database is constructed to provide diverse phase dispersion. The phase profile of the metalens is optimized by the particle swarm optimization (PSO) algorithm to minimize matching error in phase compensation. This makes it possible to obtain better matching results with less structural data, which greatly saves computing power. With the broadband linear polarization conversion ability of TNLC, the zooming of polarization-multiplexed achromatic metalens can be realized by applying different voltages. Finally, we show the electrically tunable customized dispersion-manipulated metalens and switchable color metaholograms to further demonstrate that our design is promising for spectral imaging and color display.

**Results and Discussion**

Figure 1 shows the schematic of the proposed TPAM. The main structures of the device are vertically stacked metalens and TNLCs. The building blocks of upper metalens are high-aspect-ratio birefringent titanium dioxide ($TiO_2$) nanostructures (height of 1000 nm and minimum width of 60 nm) with different cross-sectional shapes. The underlying TNLC cell is composed of two orthogonally oriented photoalignment layers with TNLC molecules in between. When there is no applied electric field, the linearly polarized light that is vertically incident and whose polarization direction is parallel to the direction of the first photoalignment layer will deflect 90° along the twisting direction of the LC molecules during the process of passing through the TNLC cell. When a voltage higher than the threshold voltage is applied to the TNLC, the long axes of the LC molecules start to incline in the direction of the electric field. When the applied voltage is large enough, except the molecules close to the photoalignment layers are anchored, the long axes of other molecules will tend to be rearranged in the direction parallel to the electric field, resulting in the disappearance of the polarization conversion of the LC molecules. Since this optical rotation of LC is independent of wavelength, the TNLC cell can work in a broadband range. Therefore, the TPAM can achieve zoom imaging by applying different voltages. The designed TPAM based on the LC electric drive has the characteristics of ultrathin and tunable focal length, which can be applied in compact and lightweight imaging devices.

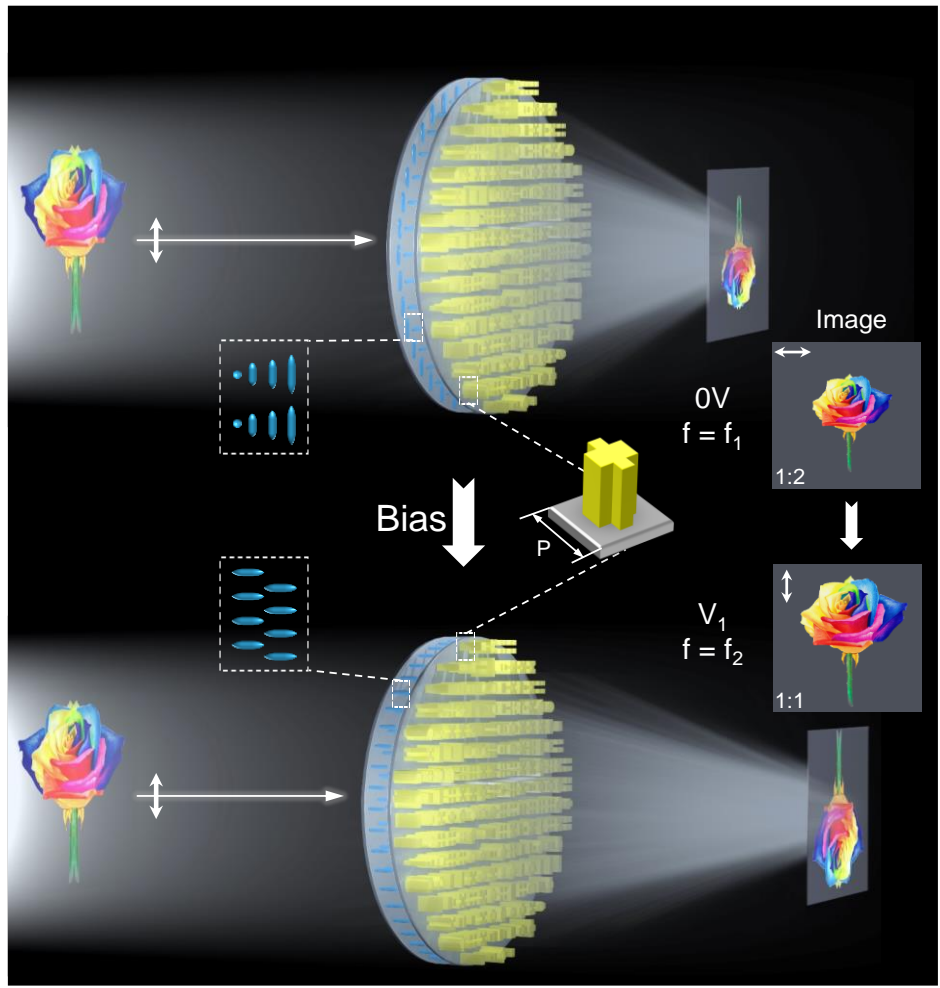

*Figure 1. Schematic of the electrically tunable polarization-multiplexed achromatic dielectric metalens (TPAM). The TPAM is integrated by twisted nematic liquid crystals (TNLCs) with all-dielectric metalens. The upper metalens is constructed of high-ratio birefringent $TiO_2$ nanostructures with different cross-sectional shapes. The underlying layer is a thin TNLC cell containing two orthogonally oriented photoalignment layers for the pre-alignment of LC molecules. By applying different voltages to the TNLC cell, the orientation of LC molecules is adjusted to realize the conversion of linearly polarized light. Achromatic zoom imaging at different voltages is experimentally demonstrated.*

To achieve polarization-multiplexed dispersion manipulation, it is necessary to design independent phase profiles at all design wavelengths under different polarizations. Each birefringent metaunit should provide a polarization-multiplexed phase response at a specific location in the metalens to satisfy the phase requirements for all design wavelengths. The phase response can be expressed as $\varphi_u = \frac{2\pi}{\lambda} n_{eff} H$ by the waveguiding effect[37], where $n_{eff}$ is the effective refractive index related with the intrinsic refractive index and cross-section shape of the metaunit, $\lambda$ is the wavelength and $H$ is the height of the metaunit. The phase response under different polarizations $\varphi_u^{x,y}$ can be modulated by adjusting and the size of metaunits parallel to the incident polarization. It can be seen that for the metaunit with a determined shape, the phase response is positively correlated with the wavenumber ($\frac{2\pi}{\lambda}$) and the exact relationship is determined by the effective refractive index. For a conventional converging spherical metalens with focal length $f$, its phase profile has to follow[38] $\varphi_{m1} = -\frac{2\pi}{\lambda}\left(\sqrt{r^2 + f^2} - f\right)$, where $\lambda$ is the design wavelength, and $r$ is the radial distance from the metalens center. Figure 2a depicts the required spatial phase profiles (phase dispersion) of the metalens for three design wavelengths. Obviously, the required phase response $\varphi_{m1}$ at selected locations is inversely related to the wavenumber, which cannot be achieved with the metaunits, as shown in the inset of Figure 2a. Therefore, the achromatic metalens schemes in the previous works[36, 39] are to shift the phase profile upward and intersect at reference position $r_0$, as shown in Figure 2b. The phase profile can be expressed as $\varphi_{m2} = -\frac{2\pi}{\lambda}\left(\sqrt{r^2 + f^2} - \sqrt{r_0^2 + f^2}\right)$. In this case, the phase dispersion transforms into a positive correlation at a location smaller than $r_0$, thus providing the possibility of phase compensation through metaunits.

However, since the introduction of $r_0$ establishes a fixed linear relationship between phase and wavenumber, it can only approximate linearly to phase dispersion of the metaunit, as shown in the inset of Figure 2b. Moreover, since the choice of metaunit database is limited, it will introduce a larger matching error for polarization-multiplexed achromatic metalens. Therefore, to achieve a better match between the dispersion of metalens and the nonlinear dispersion of metaunits, we adopted an optimized method to construct the phase profile of metalens, as shown in Figure 2c. The phase profile we constructed can be expressed as

$$\varphi_{m3} = -\frac{2\pi}{\lambda}\left(\sqrt{r^2 + f^2} - \sqrt{r_\lambda^2 + f^2}\right) \quad (1)$$

the $r_\lambda$ we chose here is a wavelength-dependent value, and the specific value can be obtained by the PSO algorithm. When the x- and y-polarization are multiplexed, the matching error between the response phase of the metaunit and the phase profile of the lens at N design wavelengths can be expressed as

$$Error = \sum_{i}^{x,y} \sum_{k=1}^{N} \left|\varphi_{m3}^i(r_\lambda^i, \lambda_k) - \varphi_u^i(\lambda_k)\right| \quad (2)$$

Through the iterative optimization of the PSO algorithm, the metaunits with the smallest matching error are found to construct a metalens. In this way, the constructed metalens phase dispersion is close to that of the metaunit, as shown in the inset of Figure 2c. Figure 2d shows the matching results of phase dispersion in the above three phase profile construction methods and metaunit phase dispersion under phase folding to 0-2π. The phase profile of each design wavelength is shifted by different amounts at the same focal point, meaning that we can construct arbitrary dispersion manipulation of the metalens to accurately match metaunit dispersion under dual-polarization channels.

The phase dispersion of metaunit is highly correlated with their cross-sectional shape, providing a new degree of freedom for dispersion manipulation of metalens. In order to obtain the required phase compensation under orthogonal linear polarization, considering the reliability and simplicity of fabrication, four types of $TiO_2$ nanostructures with rectangular composite cross-sectional were employed as the building blocks of metalens to achieve polarization sensitivity, as shown in Figure 2e. $TiO_2$ is a kind of dielectric material with a high refractive index in the visible range, which can achieve a larger range of dispersion at a lower nanometer scale. Figure 2f and 2g show the phase response database of the nanostructures with different parameters under x- and y-polarization, respectively. Details of the metaunit phase response simulation can be found in Supporting Information Section 1. Each yellow dot represents the phase response of a specifically designed nanostructure, and the red dots represent the required phase of metalens constructed according to Eq. (1), with coordinates corresponding to the folded realized phase at 450, 532 and 635 nm. Only three wavelengths are considered here to simplify the design. All of the data points form a cloud of dots that scatter across the $2\pi \times 2\pi \times 2\pi$ phase space. Folding the phase to 0-2π simplifies the phase space and avoids the one-to-one relationship of the unfolded phase. Figure 2h and 2i show the projection of the metaunit phase and the required metalens phase at coordinates 532 nm and 635 nm under x- and y-polarization, respectively. It can be seen that the required phase constructed by the previous achromatic metalens scheme cannot all find the corresponding metaunit phase to match, as shown in the upper panels of Figure 2h and 2i. This is because the constructed database is limited and cannot fill the entire phase space. After constructing the phase using Eq. (1) and optimizing $r_\lambda$ at different wavelengths using the PSO algorithm, virtually all required metalens phases can be matched with a specific metaunit, as shown in the lower panels of Figure 2h and 2i. Figure 2j and 2k show the matching results in the radius dimension for different wavelengths of the metalens with a radius of 20 μm under x- and y-polarization, respectively. The solid line is the constructed wavefront phase profiles and the dots are the phase of the matched metaunits. It can be seen that the matched metaunits can approximately realize the constructed phase of the polarization-multiplexed achromatic metalens.

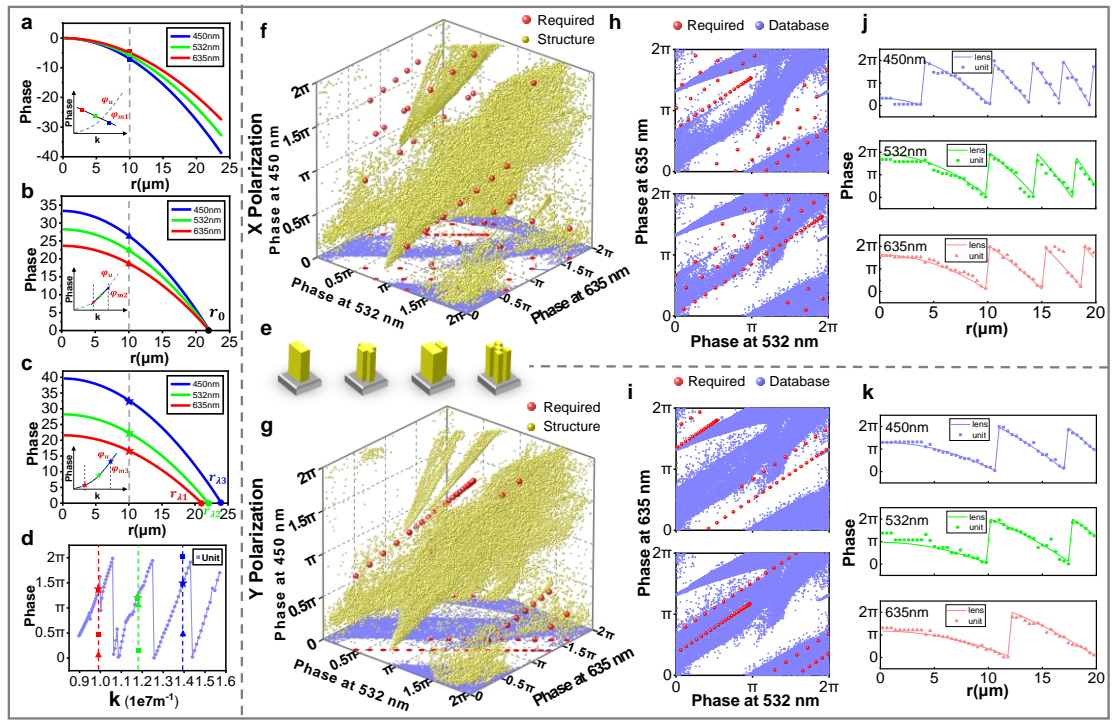

*Figure 2. Design of the TPAM. (a-c) The schematic phase profiles of conventional metalens, previous achromatic metalens scheme and optimized achromatic metalens scheme. The embedded figures are the comparison between the intrinsic phase dispersion of the metaunit*

*and constructed phase dispersion of three schemes. Three different frequencies are represented by three colors, and the phase dispersion of the three schemes are represented by different symbols. (d) The schematic matching results of phase dispersion in the above three phase profile construction schemes and metaunit phase dispersion under phase folding to 0-2π. (e) The schematic diagram of the cross-sectional of four types of $TiO_2$ nanostructures employed as the metalens components. The height of the metaunits is 1000 nm and the period is 400 nm. (f, g) The phase response database of the metaunits with different parameters under x- and y-polarization, respectively. Each yellow dot represents the phase response of a specifically designed metaunit, and the red dots represent the required phase of metalens constructed according to the previous achromatic metalens scheme, with coordinates corresponding to the folded realized phase at 450, 532 and 635 nm. (h, i) The projection of the structural phase and metalens phase at coordinates 532 nm and 635 nm under x-polarized light and y-polarized light, respectively. (j, k) The matching results for different wavelengths in the radius dimension of the metalens with a radius of 20 μm under x- and y-polarization, respectively. The solid lines represent the constructed wavefront phase profiles and the dots represent the phase of the matched metaunits.*

Figure 3a shows the detailed working principle of the TPAM. The proposed TPAM device's architecture consists of ultrathin metalens with LC molecules confined to a glass cavity. The metalens is constructed by high-aspect-ratio $TiO_2$ nanofins on a quartz substrate. The two ends of the TNLC cell are the ITO electrodes and the orthogonally oriented photoalignment layers. LC molecules are arranged in a twisted manner due to the anchoring effect of the photoalignment layer. When the input voltage is 0V, the TNLC cell converts linear y-polarization to x-polarization. When a higher voltage is input at about 5V, the TNLC cell did not have any polarization conversion. Therefore, the TPAM device produces a tunable modulation performance under the same incident light polarization conditions. To prove the concept, we prepared a sample for tunable achromatic zoom imaging. The designed electrically tunable dual-polarization channels achromatic metalens (40 μm diameter) has a focal length of 50 μm under x-polarization and 100 μm under y-polarization, and the resulting NAs are 0.371 and 0.196, respectively. The device is mainly prepared by electron-beam lithography (EBL), atomic layer deposition (ALD) and LC packaging. Details of the fabrication processes and parameters can be found in Supporting Information Section 2. The top-view and oblique-view scanning electron microscope (SEM) images of the fabricated metalens are shown in Figure 3b and 3c. Figure 3d shows the picture of the TPAM device with electrode leads. The fabricated TPAM device was measured using the optical experimental setup in Supporting Information Section 3 to characterize the focusing and imaging performance. The focal point spread functions at six wavelengths along the propagation direction (z-axis) were measured with 1 μm resolution (Figure 3e and f). The focal spot profiles and the normalized intensity profiles for each wavelength are shown in Figure 3e and f. The focal point becomes larger as the focal length increases or the wavelength increases. It can be seen that although the tunable varifocal achromatic metalens is designed at discrete three wavelengths, it can still achieve achromatic performance from 450 to 650 nm. The emergence of the secondary focus is due to some matching errors and processing errors.

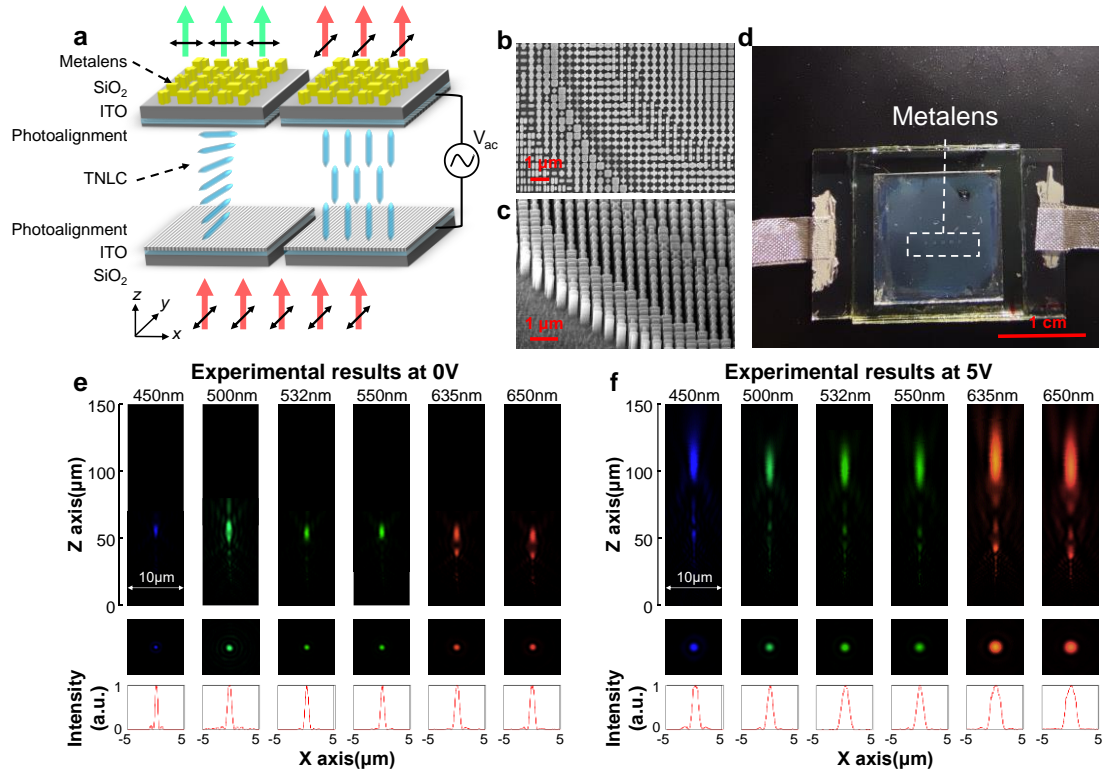

*Figure 3. Schematic diagram and experimental results of TPAM. (a) The detailed working principle of the TPAM. The LC layer here is for tunable polarization conversion enabling polarization channel selection of metalens. (b, c) The top-view and oblique-view SEM images of the fabricated dual-polarization achromatic metalens. The scale bar is 1 um. (d) The picture of the TPAM device with electrode leads. The scale bar is 1 cm. (e, f) Experimental light intensity profiles in the x-z plane, focal spot profiles and normalized intensity profiles of the TPAM device at various wavelengths.*

Figure 4a and b show the focal length and measured focusing efficiency of the TPAM device at different wavelengths under the application of different voltages. The focusing efficiency is calculated by the intensity within three times the full width half height (FWHM) at the focus plane divided by the whole intensity at the focus plane.

The orange line is the focal length of the device when 0V is applied, and the blue line is the focal length of the device when 5V is applied. The shift of the focus in the experiment is due to the processing error. It can be seen that the designed TPAM device achieves achromatic focusing while also having high focusing efficiency in the operation band. This is due to the small matching error and the high transmission of $TiO_2$ throughout the visible. To demonstrate the performance of achromatic imaging, we fabricated a dual-polarization channels chromatic metalens using the same method and compared its imaging performance with the designed TPAM device. The chromatic metalens (sample details can be found in Supporting Information Section 4) was designed for red light ($\lambda$ = 635 nm) with the same diameter and focal length as our achromatic sample. Figure 4c and d show the imaging performance of a standard 1951 United States Air Force (USAF) resolution target from the achromatic metalens and chromatic metalens illuminated by a halogen light source at various wavelengths and white light illumination at 0V and 5V, respectively. The optical experimental setup is shown in Supporting Information Section 3 and the achromatic zoom imaging process is shown in Supporting Video. In the measurements, the target and the image plane were respectively set as a fixed plane for all the wavelengths to evaluate the achromatic performance of the TPAM device. It can be seen that due to the strong chromatic effect, the line edges in the image taken with the chromatic metalens appear in multiple colors, resulting in blurred image features. The color difference is mainly from the efficiency variation of the fabricated chromatic metalens. In contrast, the image taken with the TPAM device exhibits clear line features, showing the chromatic effect is completely eliminated.

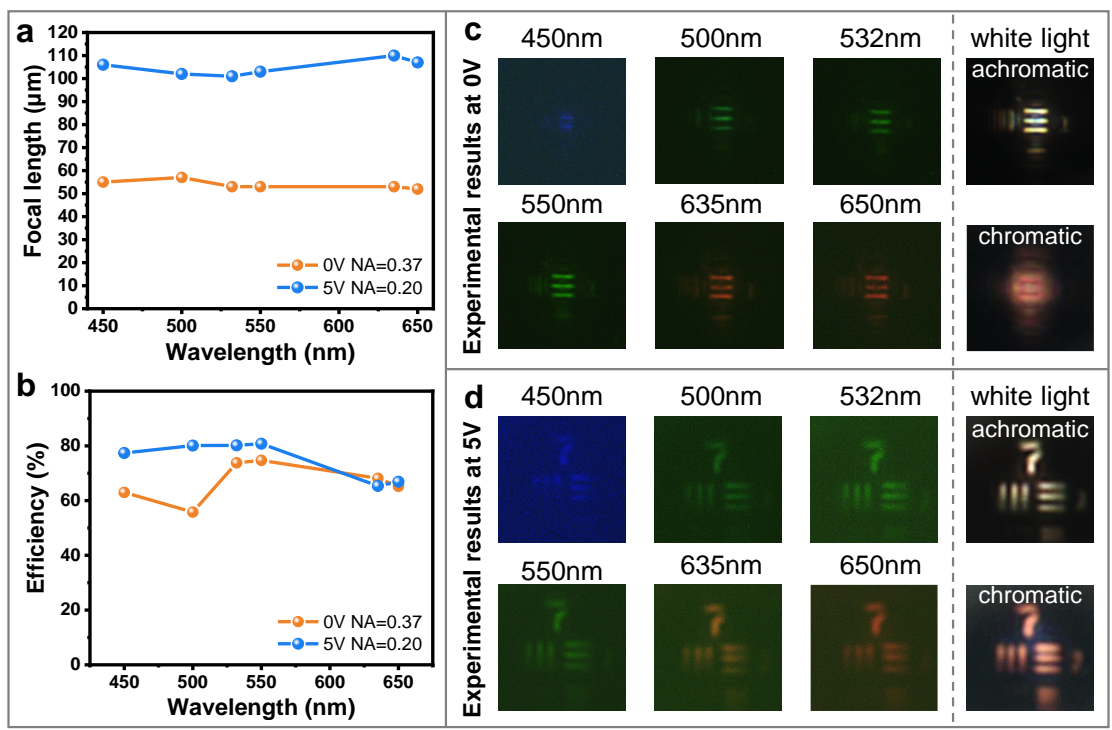

*Figure 4. Focus characterization and imaging performance of TPAM device. (a) Focal length distribution at different wavelengths. The orange line is the focal length of the device when 0V is applied, and the blue line is the focal length of the device when 5V is applied. (b) The measured focusing efficiency of the TPAM devices. The orange line is the focusing efficiency of the device when 0V is applied, and the blue line is the focusing efficiency of the device when 5V is applied. (c, d) Images of element 1 in group 7 on the 1951 United States Air Force resolution target formed by the TPAM device at various wavelengths and white light.*

In order to further demonstrate the practicality of the optimized metalens scheme for dispersion manipulation, we designed and fabricated a metalens device with customized dispersion relationships. When y-polarized light is incident, this device exhibits a positive dispersion relationship at 0V and a negative dispersion relationship at 5V. Customized dispersion design is achieved by constructing the wavefront phase at each wavelength for the desired focus, where the optimized metalens scheme provides great freedom to design each wavefront. The designed customized dispersion-manipulated metalens has a diameter of 40 μm and a focal length at 0V of 60 μm at 450 nm, 80 μm at 532 nm, 100 μm at 635 nm and a focal length at 5V of 100 μm at 450 nm, 80 μm at 532 nm, 60 μm at 635 nm. Figure 5a and b show the measured intensity distribution along the propagation direction, focal spot profiles and normalized intensity profiles for the customized dispersion-manipulated metalens at three incident wavelengths under different voltages. Figure 5c shows the measured focal length at designed wavelengths. The orange line is the focal length of the positive dispersion when 0V is applied, and the blue line is the focal length of the negative dispersion when 5V is applied. The black line is the negative dispersion reference curve for the measured chromatic metalens. It can be seen that the designed metalens achieve good focus at each wavelength with significantly different chromatic aberrations. Further, we demonstrated the electrically switchable dual-polarization channels color metaholograms operating at 450, 532, and 635 nm with the method of phase matching. The phase profiles of the target image at different wavelengths were retrieved by the Gerchberg-Saxton algorithm. The target images were well reconstructed and switched at different voltages, although the contrast is reduced due to phase matching errors and processing errors.

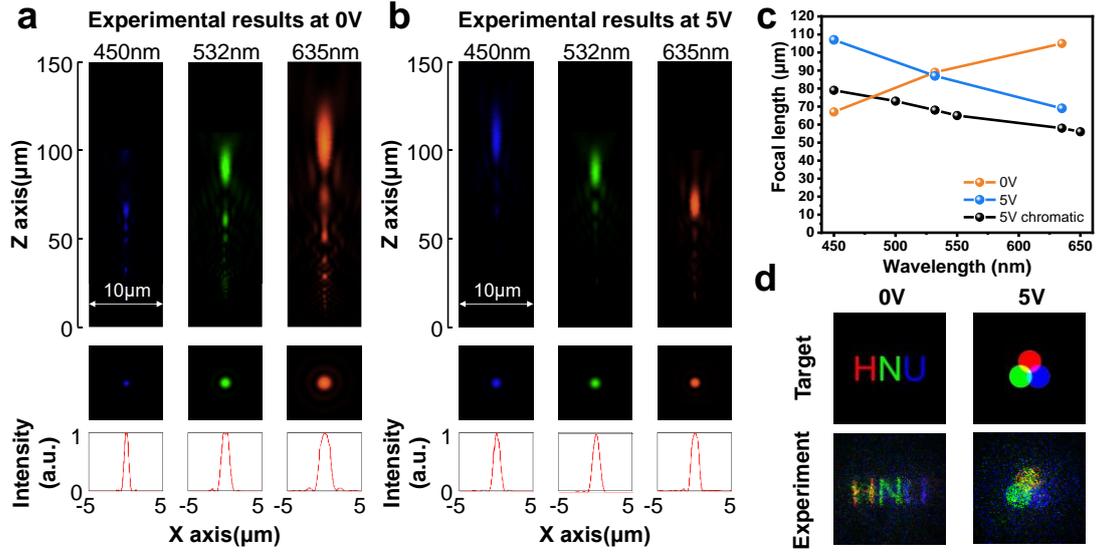

*Figure 5. Experimental results of customized dispersion-manipulated metalens and color metaholograms.* *(a, b) Experimental light intensity profiles, focal spot profiles and normalized intensity profiles for the customized dispersion-manipulated metalens at designed incident wavelengths. (c) Focal length distribution at different wavelengths. The orange line is the focal length of the positive dispersion metalens when 0V is applied, and the blue line is the focal length of the negative dispersion metalens when 5V is applied. The black line is the negative dispersion reference curve for the measured dual-polarization channels chromatic metalens designed at 635 nm when 5V is applied. (d) Target image and the measured color metaholograms designed at 450, 532, and 635 nm.*

The proposed dispersion manipulation scheme can use optimization algorithms such as the PSO algorithm and genetic algorithm to optimize the phase profile of each wavelength under different polarization. This makes it possible to obtain better matching results with less structural data. The number of designed discrete wavelengths can be increased to obtain better achromatic results, but this will increase computational complexity. The customized dispersion manipulation capability can be applied in various applications such as AR/VR displays and spectral detection. A continuous zoom metalens can also be achieved through a more refined design[40]. Due to the large degree of freedom of metasurfaces in polarization control, combined with the polarization control function of liquid crystals, more tunable devices can be realized[41, 42].

**Conclusion**

In summary, we proposed a tunable broadband metalens design strategy in the visible region. The vertically stacked integration of metalens and TNLCs provides a simpler device fabrication process and control mechanism. The phase profiles of the achromatic metalens at each designed wavelength are optimized by the PSO algorithm to minimize matching errors in phase compensation. With this scheme, the polarization-multiplexed achromatic metalens composed of a series of $TiO_2$ nanostructures with a high aspect ratio achieves zoom imaging in the range of 450nm to 650nm. Further, the electrically customized dispersion-manipulated metalens and switchable color metaholograms are demonstrated. The proposed scheme provides a great degree of freedom to realize various types of customized dispersion modulation and tunable modulation for applications such as AR/VR displays and spectral imaging. In addition, it provides a new idea for the realization of tunable broadband metasurfaces.

**Acknowledgments**

The authors acknowledge the financial support from, the National Natural Science Foundation of China (Grant No. 52005175, 52111530233), the National Key

Research and Development Program of China (Grant No. 2021YFB3600500), Natural Science Foundation of Hunan Province of China (Grant No. 2022JJ20020), Shenzhen Science and Technology Program (Grant No. RCBS20200714114855118) and the Tribology Science Fund of State Key Laboratory of Tribology (SKLTKF20B04).

**Author Contributions:**

Y.H., H.D. and X.O. proposed the idea. Y.H., X.O. and Y.Z. conceived and carried out the design and simulation. X.O., Y.J. and Z.G. prepared the metasurface samples. T.Z. and F.F. carried out the LCs packaging. Y.H., X.O. and T.Z. conceived and performed the measurements. Y.H., X.O., T.Z., F.F. and H.D. analyzed the results and co-wrote the manuscript. All the authors discussed the results and commented on the manuscript.

**Competing Interests:** The authors declare that they have no competing interests.